\documentclass[aps,twocolumn,showpacs]{revtex4}
\usepackage{graphicx}
\usepackage{bm}
\usepackage{amsmath}

\begin{document}

\title{Light Induced Hall effect in semiconductors with spin-orbit coupling}
\author{Xi Dai$^{1,2}$, and Fu-Chun Zhang$^{1} $}
\affiliation{$^1$ Department of Physics, and Center of Theoretical and Computational
Physics, the University of Hong Kong, Hong Kong\\
$^2$ Institute of Physics, Chinese Academy of Sciences, Beijing, China}
\date{\today}

\begin{abstract}
We show that optically excited electrons by a circularly polarized light in
a semiconductor with spin-orbit coupling subject to a weak electric field
will carry a Hall current transverse to the electric field. This light
induced Hall effect is a result of quantum interference of the light and the
electric field, and can be viewed as a physical consequence of the spin
current induced by the electric field. The light induced Hall conductance is
calculated for the p-type GaAs bulk material, and the n-type and p-type
quantum well structures.
\end{abstract}

\pacs{72.15Gd,73.63.Hs,75.47.-m,72.25.-b}
\maketitle

The phase coherent semiconductor spintronic device is an important candidate
for the quantum devices, which allows the storage, manipulation and
transport of quantum information\cite{review1}. Due to the quantum nature of
a spin system, a single electron with spin 1/2 is an ideal qbit for quantum
computing and an ideal unit for data storage. Therefore, the study of spin
transport is very important for the future development of spintronic
techniques. One of the most common methods in manipulating and detecting an
electron's spin state is the optical absorption or emission of the
circularly polarized light (CPL)\cite{review2}. The spin polarized charge
current may be induced by the absorption of the CPL, which is called
circular photogalvanic effect(CPGE)\cite{CPGE}. The CPGE was first proposed
almost thirty years ago\cite{CPGE1} and has been detected in both bulk
materials and semiconductor quantum well(QW) structures\cite{CPGE2}.

In this Letter, we propose a new effect for a broad class of semiconductors
with spin-orbit coupling, which we shall call light induced Hall effect
(LIHE). In that effect, optically excited electrons by a CPL in the
semiconductor subject to a weak static electric field will carry a Hall
current transverse to the electric field. Different from the CPGE, which
only occurs in materials with inversion asymmetry, the LIHE occurs also in
bulk zinc-blende structure materials such as GaAs where the inversion
symmetry is preserved\cite{review1}. The LIHE may be viewed as a response of
the local spin Hall current induced by the electric field in the spin-orbit
system to the CPL\cite%
{SHE1,SHE2,hu,shen,shenprl,inoue,rashba,review3,mishchenko,dai}. In the
materials with the structure inversion asymmetry the LIHE is expected to be
more pronounced in the case where the incident light is normal to the sample
to eliminate the CPGE\cite{review2}. In particular, we predict and estimate
the effect by calculating the Hall photocurrent for three different systems:
the p-type GaAs bulk material, the n-type and p-type GaAs quantum well
structures.

\begin{figure}[tbp]
\begin{center}
\includegraphics[width=6cm,angle=0,clip=]{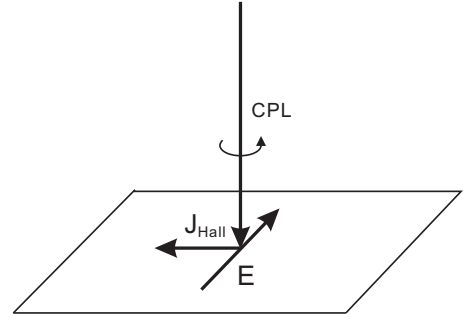}
\end{center}
\caption{The Hall photocurrent generated by the circularly polarized light
and static electric field. }
\label{fig1}
\end{figure}

We begin with a more detailed description of and general discussions on the
LIHE, followed by explicit calculations of the effect on prototype systems.
Let us consider a semiconductor with an incident CPL along the z-axis with $%
\vec{e_p}$ its Poynting unit vector, and a weak external static electric
field $\vec{E}$ along the x-axis. Similar to the ordinary Hall effect, a
transverse electric current along the y-direction will be generated in
addition to the current along the x-direction. The schematic plot of the
LIHE can be simply illustrated in fig\ref{fig1}. The transverse current in
this case is entirely induced by the CPL through the optical transition from
the valence to the conduction band and its direction and magnitude can be
determined by $\mathbf{J}_{hall}=\sigma _{xy}\lambda \mathbf{E\times e}_{P}$
, where $\sigma _{xy}$ is the light induced Hall conductivity, $\lambda =\pm
1$ is the helicity of the CPL. From the symmetry point of view, the CPL in
the LIHE plays the similar role as the magnetic field in the ordinary Hall
effect to break the time reversal symmetry. However, unlike the ordinary
Hall effect, the LIHE is purely a quantum effect induced by the spin-orbit
coupling. As we will show below, the LIHE is induced by the Berry curvature
of the band structure in the k-space.

LIHE may be understood as a quantum interference effect between the CPL and
the static electric field. As discussed by Murakami et al.\cite{SHE1} and by
Sinova et al.\cite{SHE2}, when an electron (or a hole) moving along the
y-direction is accelerated along the $x$-direction due to the electric
field, its spin will tilt upward or downward along the z-direction. The
electrons (or holes) moving with opposite momentum along the y-direction in
the electric field will tilt their spins with one upward and the other
downward, thus generating a non-zero spin current $j_y^z = 1/2(v_y\sigma^z +
\sigma^z v_y)$. This spin Hall effect has generated a lot of research
interest recently\cite%
{SHE1,SHE2,hu,shen,shenprl,inoue,rashba,review3,mishchenko,dai}. In the
presence of the right handed CPL, in the GaAs bulk or quantum well, the
electrons will be pumped from the valence to the conduction band. Within the
dipole approximation, only an electron with total angular momentum along the
$z$-axis $J_{z}=-3/2$ ($J_{z}=-1/2$) will absorb the CPL and jump to the
conduction band with $S_{z}=-1/2$ ($S_{z}=1/2$). Therefore if the electron
spin in the valence band tilt upwards or downwards, the corresponding
transition rate to the conduction band will then be enhanced or suppressed
due to the transition selection rule. As a consequence, the imbalance of the
photo-excited electron density of the conduction band in the $k$ space along
the $y$-direction will also be induced, which leads to a spin polarized
current along the $y$ direction.

LIHE can also be viewed as the optical response of the system carrying a
pure spin current. The possible physical consequence induced by the spin
current is a highly interesting issue in the field of spintronics. The LIHE
generated by the spin current can then be either used to detect the
existence of a spin current or to design the new type of quantum devices.
The spin current generated by spin Hall effect is very difficult to detect.
Up to now, the only way to measure the spin current flowing through the
sample is to measure the spin accumulation at the edges generated by the
spin current\cite{sih,wunderlich}. Although the spin accumulation has been
detected by two experiments using Kerr effect and photo luminescence spectra
respectively, the quantitative relationship between the spin accumulation at
the sample edge and the strength of the spin current flowing through the
bulk is still not very clear. Since the measurement of the charge current
can be relatively easily carried out by detecting the magnetic field built
up around the current, for instance, the LIHE should shed light on the new
methods of measuring spin current.

In what follows, we will discuss the LIHE in three different systems, namely
the 3D hole system described by the Luttinger model, the 2D hole gas
described by the Luttinger model under the confinement potential along the
z-direction and the 2D electron gas described by the Rashba model.

We first consider a single particle Hamiltonian with momentum $\vec P$ in
the bulk GaAs\cite{winkler},
\begin{eqnarray}
H_{v}(\overrightarrow{P}) &=&\frac{P^{2}}{2m}(\gamma _{1}+\frac{5\gamma _{2}%
}{2})-\frac{\gamma _{2}}{m}(\vec{S}\cdot \overrightarrow{P})^{2}  \label{H_v}
\\
H_{c}(\overrightarrow{P}) &=&\frac{P^{2}}{2m_{e}}  \label{H_c}
\end{eqnarray}
for a hole in the valence band and an electron in conduction band,
respectively.

The above Hamiltonian can be easily diagonalized. To calculate the
modification of the inter-band transition rate induced by the static
electric field, we use a non-linear response theory, where the second order
correction combining the electric field $\overrightarrow{E}$ \ and the
intensity of the light $I$ will be taken into account. This high order
response term can be obtained by the following way. First we switch off the
light field and obtain the approximate wave function to the first order of
the static electric field $\overrightarrow{E}$. Then we switch on the light
field and use the $\overrightarrow{E}$ \ dependent wave function to
calculate the transition rate. Following references\cite{hu,fradkin}, the
electric field is included in the Hamiltonian through the vector potential $%
\overrightarrow{A}=$ $\overrightarrow{E}t$ and the momentum $\overrightarrow{%
P}$ in equation \ref{H_c} and \ref{H_v} is replaced by $\overrightarrow{P}-e%
\overrightarrow{E}t$.

.

We assume the electric field is switched on at time $t=0$ and obtain the
first-order time-dependent wave function $\left\vert m,k,t\right\rangle ^{E}$
for such a system in terms of the instantaneous eigenstates,

\begin{gather}
\left\vert m,\mathbf{k},t\right\rangle ^{E}=\exp \left\{
-i\int_{0}^{t}dt^{\prime }\varepsilon _{m}(\mathbf{k},t^{\prime })/\hbar
\right\} \{\left\vert m,\mathbf{k},t\right\rangle +i\sum_{n\neq m}  \notag \\
\frac{\left\vert n,\mathbf{k},t\right\rangle \left( f_{n,k}-f_{m,k}\right)
\mathbf{\Omega }_{nm}\left( \mathbf{k,}t\right) \bullet e\mathbf{E}}{\left[
\varepsilon _{n}(\mathbf{k},t)-\varepsilon _{m}(\mathbf{k},t)\right] }\left[
1-e^{-i\left( \varepsilon _{n}(t)-\varepsilon _{m}(t)\right) t/\hbar }\right]
\}  \label{wf1}
\end{gather}

where $\left\vert m,k,t\right\rangle $ is the instantaneous eigenstates of
the Hamiltonian $H_{v}(\hbar \overrightarrow{k}-e\overrightarrow{E}t)$,
which satisfies

\begin{equation}
H_{v}(\hbar \mathbf{k}-e\mathbf{E}t)\left\vert n,\mathbf{k},t\right\rangle
=\varepsilon _{n}(t)\left\vert n,\mathbf{k},t\right\rangle  \label{Ht}
\end{equation}%
, $f_{n,k}$ is the Fermi distribution function and $\Omega _{nm}\left(
\mathbf{k,}t\right) =\left\langle n,\mathbf{k},t\right\vert \frac{\partial }{%
\partial \mathbf{k}}\left\vert m,\mathbf{k},t\right\rangle $ is the Berry
curvature of the Bloch states.

We then switch on the CPL. The optical transition rate can be obtained by
solving the time dependent Schrodinger equation perturbatively. After
lengthy derivation, we can prove that the optical transition rate in the
presence of the electric field can be obtained by simply using the above
wavefunction in the Fermi golden rule, which reads

\begin{gather}
\Gamma _{mk,\alpha }\approx \{\left\vert \left\langle \alpha ,\mathbf{k}%
\right\vert W\left\vert m,\mathbf{k}\right\rangle \right\vert
^{2}+2\sum_{n\neq m}  \notag \\
\mathbf{Re}[i\frac{W_{\alpha n}(\mathbf{k})\mathbf{\Omega }_{nm}\left(
\mathbf{k,}0\right) W_{m\alpha }(\mathbf{k})\bullet e\mathbf{E}\left(
f_{n,k}-f_{m,k}\right) }{\left[ \varepsilon _{n}(\mathbf{k})-\varepsilon
_{m}(\mathbf{k})\right] }]\}  \notag \\
\times f_{m,k}\left( 1-f_{\alpha ,k}\right) \delta \left( \hbar \omega
-\varepsilon _{\alpha }(\mathbf{k})+\varepsilon _{m}(\mathbf{k})\right)
\label{rate}
\end{gather}

The matrix $\overset{\wedge }{W}$ describes the coupling between the
electrons in the solid and the right handed CPL in the dipole approximation
and takes the form of
\begin{equation*}
\overset{\wedge }{W}=\left(
\begin{array}{ccccc}
& 3/2 & -3/2 & 1/2 & -1/2 \\
1/2 & g & 0 & 0 & 0 \\
-1/2 & 0 & 0 & \frac{g}{\sqrt{3}} & 0%
\end{array}%
\right)
\end{equation*}

where $g=dE_{rad}$ with $d$ the effective dipole induced by the CPL and $%
E_{rad}$ the amplitude of the electric field of the CPL. Assuming the power
density of the CPL to be $100mW/mm^{2}$ and $d= 4.8\times 10^{-29}C\cdot m$%
\cite{yu}, we estimate the coupling energy of electron and CPL to be $%
2.\,\allowbreak 603\,8\times 10^{-6}eV$.

Within the simplest relaxation approximation, we can express the Hall
photocurrent as the summation of the electron and hole currents,

\begin{equation}
\left\langle \overrightarrow{j}_{total}\right\rangle \mathbf{=}\sum_{\mathbf{%
k}}\sum_{m\alpha }\left[ e\overrightarrow{v}_{\alpha \alpha }^{e}\left(
\mathbf{k}\right) \Gamma _{mk,\alpha }\tau _{e}-e\overrightarrow{v}%
_{mm}^{h}\left( \mathbf{k}\right) \Gamma _{mk,\alpha }\tau _{h}\right]
\end{equation}

where $\overrightarrow{v}_{k}^{e}$ and $\overrightarrow{v}_{k}^{h}$ are the
velocity operator, $\tau _{e}$ and $\tau _{h}$ are the relaxation time for
the electrons and holes respectively. For circularly polarized light
propagate perpendicular to the xy-plane, the contribution from the first
term in equation \ref{rate} cancels exactly after integrating over k.
Therefore in the present case the total charge current,which is found along
the y direction, is purely induced by the static external electric field.
Similar to the Hall effect, we can express the transverse charge current in
terms of the electric field, which reads $J_{y}=\sigma _{xy}^{ph}E_{x}$ with
\bigskip

\bigskip

\begin{gather}
\sigma _{xy}^{ph}=\sum_{m\alpha }\sum_{\mathbf{k}}\left[ ev_{\alpha \alpha
,x}^{e}\left( \mathbf{k}\right) \Gamma _{mk,\alpha }\tau
_{e}-ev_{mm,x}^{h}\left( \mathbf{k}\right) \Gamma _{mk,\alpha }\tau _{h}%
\right] \times  \notag \\
\{2\sum_{n\neq m}\mathbf{Re}[i\frac{W_{\alpha n}(\mathbf{k})\mathbf{\Omega }%
_{nm}\left( \mathbf{k,}0\right) W_{m\alpha }(\mathbf{k})\cdot \mathbf{e}%
_{y}\left( f_{n,k}-f_{m,k}\right) }{\left[ \varepsilon _{n}(\mathbf{k}%
)-\varepsilon _{m}(\mathbf{k})\right] }  \notag \\
\times f_{m,k}\left( 1-f_{\alpha ,k}\right) \delta \left( \hbar \omega
-\varepsilon _{\alpha }(\mathbf{k})+\varepsilon _{m}(\mathbf{k})\right) \}
\label{sig_xy}
\end{gather}

For the 3D Luttinger model, we can solve the unperturbated Hamiltonian
analytically and obtain a very simple analytical expression for the light
induced Hall conductance in the low temperature as

\begin{equation}
\sigma _{xy}^{ph}=\sum_{m=L,H}\frac{3\pi ^{2}+2}{16\hbar \omega m_{c}\pi }%
\frac{\alpha Ie^{2}\tau m_{0}\hbar \left( f_{\overset{\_}{m}%
,k_{m}}-f_{m,k_{m}}\right) }{\left( \hbar \omega -E_{g}\right) \gamma
_{2}\mu _{m}}  \notag
\end{equation}

where $\alpha $ is the optical absorption coefficient, which is around $%
10^{4}cm^{-1}$for GaAs, $I$ is the intensity of the light, $\hbar \omega $
is the energy of the photo, $E_{g}$ is the energy gap of GaAs,$k_{m}$
satisfies $\hbar \omega -\varepsilon _{c}(\mathbf{k}_{m})+\varepsilon _{m}(%
\mathbf{k}_{m})=0$, $\tau $ is the momentum relaxation time for the
electrons in the conduction band, $m_{0}$ is the bare electron mass, $m_{c}$
is the effective mass for the conduction band and $\mu _{m}$ is the optical
effective mass defined as $\mu _{m}^{-1}=m_{c}^{-1}+m_{v}^{-1}$. If we
choose the typical experimental parameters for GaAs as $\alpha
=10^{4}cm^{-1},\tau _{e}=10^{-12}s,I=100mW/mm^{2},E_{g}=1.42eV,\hbar \omega
=1.67eV,$ we can obtain the value of the light induced Hall conductance to
be $7.\,\allowbreak 580\,5\times 10^{-3}\Omega ^{-1}m^{-1}.$

We can also calculate the light induced Hall conductance defined in equation %
\ref{sig_xy} for the quantum well structure as well. In this case, the
applied CPL must be normal to the plane. Without the external electric
field, the CPL can only induce a pure spin current within the plane\cite%
{sipe,dai2}. The only difference here is using the eigen states for the
subbands in the above equation \ref{sig_xy}. In the present study, we
calculate the light induced Hall conductance for both p-type and n-type
quantum well samples numerically. \ The Hamiltonian of the GaAs quantum well
structure can be written as

\begin{eqnarray}
H_{v,well}(\overrightarrow{P}) &=&H_{v}(\overrightarrow{P})+V(z)+\lambda
_{v}\left( \overrightarrow{P}\times \overrightarrow{S}\right)  \label{H_vw}
\\
H_{c,well}(\overrightarrow{P}) &=&H_{c}(\overrightarrow{P})+V(z)+\lambda
_{c}\left( \overrightarrow{P}\times \overrightarrow{\sigma }\right)
\label{H_cw}
\end{eqnarray}

\begin{figure}[tbp]
\begin{center}
\includegraphics[width=7cm,angle=0,clip=]{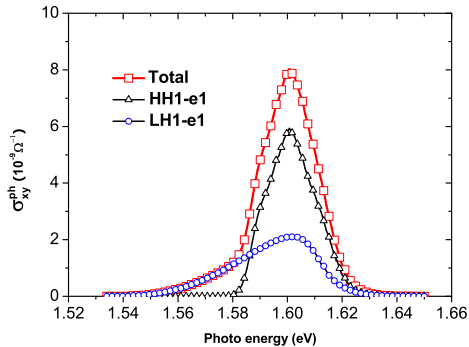}
\end{center}
\caption{The light induced Hall conductance in a p-type GaAs quantum well as
the function of photon energy. }
\label{fig2}
\end{figure}

\begin{figure}[tbp]
\begin{center}
\includegraphics[width=7cm,angle=0,clip=]{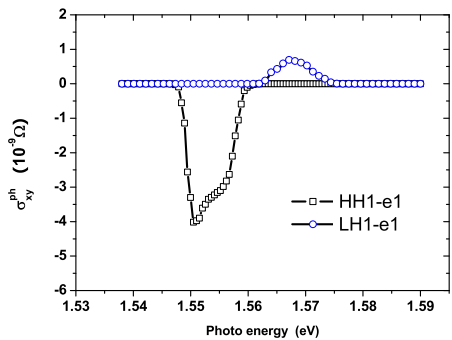}
\end{center}
\caption{The light induced Hall conductance in a n-type GaAs quantum well as
the function of photon energy. }
\label{fig3}
\end{figure}

where $H_{v}(\overrightarrow{P})$ and $H_{c}(\overrightarrow{P})$ are the
Hamiltonian for the holes in the valence band and electrons in the
conduction band, $\lambda _{v}$ and $\lambda _{c}$ are the effective Rashba
coupling \ for the valence band and conduction band respectively which are
induced by the structure inversion symmetry breaking, $\ $and $%
\overrightarrow{S}$ , $\overrightarrow{\sigma }$ are the spin 3/2 and 1/2
matrices for valence band and conduction band respectively. In the present
paper, we choose the confinement potential $V(z)=+\infty $ for $|z|>L$ and $%
V(z)=0$ otherwise.\bigskip\ Using the numerical techniques presented in
detail in our previous paper\cite{dai}, we first obtain the subband
dispersions for the e1, HH1 and LH1 subbands, which are plotted in Fig\ref%
{fig1}. In the calculation, we choose $\gamma _{1}=7.0$, $\gamma _{2}=1.9$, $%
m_{e}=0.067m_{0}$, where $m_{0}$ is the bare electron mass. Then we
calculate the light induced Hall conductance for both the p-type and n-type
quantum well structure with the following parameters, $\tau
_{e}=5.\,\allowbreak 886\,3\times 10^{-11}s$, $I=100mW/mm^{2}$, $%
g=2.\,\allowbreak 603\,8\times 10^{-6}eV$. The carrier density is chosen to
be $9.\,\allowbreak 280\,7\times 10^{10}cm^{-2}$ for the n-type case and $%
2.\,\allowbreak 626\,1\times 10^{11}cm^{-2}$ for the p-type case. The
results are shown in Fig\ref{fig2} and \ref{fig3} respectively. In the
present study, we only include the transition between the HH1,LH1 and e1
subbands. The difference behavior of the light induced Hall effect between
the n-type and p-type samples quite clear in Fig\ref{fig2} and \ref{fig3}.
In the n-type sample the contribution to the light induced Hall conductance
from the $HH1-e1$ transition has a different sign with that of the $HH1-e1$
transition. While in the p-type sample, the contribution comes from two
different transitions have the same sign. This interesting asymmetric
behavior of n-type and p-type samples can be understood in the following
way. In the n-type sample, the FS lies within the subband $e1$ and the spin
of the electrons will tilt out of the plane when an electric field is
applied. Suppose we use the right hand CPL here. According to the selection
rule, for the HH1-e1 transition the only allowed process is from $%
|S_{z}=-3/2>$ in the valence band to $|s_{z}=-1/2>$ in the conduction band.
And that of the LH1-e1 transition is from $|S_{z}=-1/2>$ in the valence band
to $|s_{z}=1/2>$ in the conduction band. Thus when the electron spin tilt
out of the plane, the induced modification of the two transition rate will
be opposite in sign , which gives the opposite sign for the light induced
Hall current.

From equation \ref{sig_xy}, we know that the strength of the LIHE is
determined by the optical coupling matrix and the Berry curvature of the
Bloch states at the manifold in the k-space which satisfy the energy
conservation. The physical consequence of the Berry curvature in K-space was
first found in the Anomalous Hall effect and late in the spin Hall effect.
Then the LIHE we proposed here can be also viewed as a new physics
consequence of the Berry curvature in K-space.

Another important issue we would like to discuss is the role of the disorder
in the LIHE. As discussed in reference \cite{inoue} and \cite{mishchenko},
if the spin orbital coupling in the system has the linear dependence in k,
the disorder effect will exactly cancel the spin Hall effect in the thermal
dynamic limit through the vertex correction terms. Therefore in such
systems, the spin current induced by spin Hall effect only exists in the
mesoscopic scale and vanishes in the macroscopic scale. Since the LIHE is
generated by the direct optical absorption modulated by the static electric
field, to see the LIHE it only requires the spin current to be generated in
the scale of the light wave length, which is in the mesoscopic scale for the
GaAs. Therefore, unlike the spin Hall effect, for such kind of systems, i.e.
the n-type GaAs quantum well structure described by the Rashba model, the
LIHE can also survive even for the macroscopic samples.

In summary, we have proposed a new effect, Light induced Hall effect in this
paper. This effect is generated by the modulation of the optical transition
rate in the k-space induced by the static electric field. The LIHE can be
viewed in several different ways. First it can be viewed as the quantum
interference effect between two different external fields, the light field
and the static electric field. Secondly, the LIHE can be thought as the
physics consequence generated by the non-zero spin current flowing through
the bulk generated by the spin Hall effect.Thirdly, the LIHE can also be
viewed as the physical effect reflecting the Berry curvature of the Bloch
state in the k-space. We have also calculated the Light induced Hall
conductance for three different semiconductor systems and made the
quantitative predictions.

\end{document}